\documentclass[aps,prb,twocolumn,superscriptaddress,showpacs]{revtex4}
\usepackage{graphicx}
\usepackage{amsmath}
\usepackage{amssymb}

\newcommand{\dmo}[4]{{#1}$_{#2}${#3}$_{#4}$MnO$_3$}

\graphicspath{{.}}

\begin{document}

\title{Mixed-phase description of colossal magnetoresistive manganites}

\author{Alexander Wei{\ss}e}
\affiliation{Physikalisches Institut, Universit\"at Bayreuth, 95440
  Bayreuth, Germany}

\author{Jan Loos}
\affiliation{Institute of Physics, Czech Academy of Sciences, 16200
Prague, Czech Republic}

\author{Holger Fehske}
\affiliation{Institut f\"ur Physik, Ernst-Moritz-Arndt Universit\"at
  Greifswald, 17487 Greifswald, Germany}
\date{\today}

\begin{abstract}
  In view of recent experiments, indicating the spatial coexistence of
  conducting and insulating regions in the ferromagnetic metallic
  phase of doped manganites, we propose a refined mixed-phase
  description. The model is based on the competition of a
  double-exchange driven metallic component and a polaronic insulating
  component, whose volume fractions and carrier concentrations are
  determined self-consistently by requiring equal pressure and chemical
  potential. The resulting phase diagram as well as the order of the
  phase transition are in very good agreement with measured data. In
  addition, modelling the resistivity of the mixed, percolative phase
  by a random resistor network, we obtain a pronounced negative
  magnetoresistance in the vicinity of the Curie temperature $T_C$.
\end{abstract}

\maketitle

\section{Motivation}

The peculiar properties of the ferromagnetic metallic phase of
mixed-valence manganites (e.g. \dmo{La}{1-x}{Ca}{x} with $0.2\lesssim
x\le 0.5$), in particular the large negative magnetoresistance close
to the Curie temperature $T_C$, have been the subject of intense
research activity over the last years~\cite{Ra97,CVM99,TT99,DHM01}.
There is now compelling experimental evidence, that the complex
interplay of the electronic degrees of freedom (charge, spin and
orbital) and the lattice leads to the spatial coexistence of regions
with different properties rather than to the formation of a single
homogeneous phase.  Above $T_C$ the activated behaviour of the
conductivity~\cite{JSRTHC96,PRCZSZ97,WMG98}, the Hall
conduction~\cite{JHSRDE97} and the
thermopower~\cite{JSRTHC96,PRCZSZ97}, as well as the structure of the
pair distribution function (PDF)~\cite{BDKNT96} indicate the formation
of small polarons, i.e., of almost localised carriers within a
surrounding lattice distortion. Interestingly these polarons continue
to exist in the metallic phase below $T_C$, merely their volume
fraction is noticeable reduced. Corresponding evidence is found in
conductivity measurements~\cite{Jaea99,ZSPK00}, muon spin
relaxation~\cite{Heea00,Heea01}, high-resolution x-ray
diffraction~\cite{BPPSK00}, pulsed neutron diffraction~\cite{LEBRB97}
and in x-ray absorption fine structure spectra~\cite{LSBNBRC98}.
Based on these experimental observations different scenarios for the
coexistence of conducting and insulating regions within the metallic
phase of the manganites were discussed, which relate the
metal-insulator transition to phase separation~\cite{DHM01} and
percolative phenomena~\cite{GK98,GK99,GK00,MMFYD00}. In particular
microscopic imaging techniques, like scanning tunneling
spectroscopy~\cite{FFMTAM99,Beea02} or dark-field
imaging~\cite{UMCC99}, seem to support the latter idea. However, as
yet the detailed nature of the spatially coexisting regions or phases
is not known very precisely and even the data for the corresponding
length scales is contradictory~\cite{Raea01,Siea02}.

In a recent work~\cite{WLF01b} we addressed the problem of coexisting
conducting and insulating regions within the metallic phase of the
manganites and proposed a phenomenological mixed-phase description,
which is based on the competition of a polaronic insulating phase and
a metallic, double-exchange driven ferromagnetic phase. Both phases
are assumed to have an equal density of charge carriers and the
percolative coexistence is accounted for by a metallic bandwidth,
which depends on the volume fractions of the two components. The model
is able to describe a finite polaronic volume fraction well below
$T_C$ and yields rather realistic $x$-$T$-phase diagrams. However, its
sensitivity to external magnetic fields is much too weak and we made
no attempt to describe resistivities or the large magnetoresistance.

Due to the insufficient knowledge of the two different components of
the low temperature phase, the assumptions of our previous model may
be problematic, in particular the balance of the two components could
follow from different equilibrium conditions.  In the present work we
discard the condition of equal charge density within the polaronic
insulating and the ferromagnetic metallic regions and require equal
pressure instead. This approach is well justified, since one of the
components is insulating and the length scale of the coexistence seems
to be short enough, as to avoid long range Coulomb effects. We
complete this new mixed-phase description by a model for the resistivity
which is based on a random resistor network that accounts for the
percolative nature of the low temperature phase.  The phase diagram we
obtain from the improved model is comparable to the previous results,
however, the sensitivity to external magnetic fields is much stronger
and the ansatz for the conductivity yields a rather large
magnetoresistance close to $T_C$.  In addition, the order of the phase
transition from the ferromagnetic metallic to the paramagnetic
insulating phase depends on the model parameters and in particular on
the doping $x$, a feature which was observed for the real
materials~\cite{MRRVL99,Kiea02}.

\section{Coexisting components}
In the doping range $0.2\lesssim x\le 0.5$ the electronic properties
of the manganites are dominated by the well known double-exchange
interaction~\cite{Ze51b,AH55,KO72a} and the electron-lattice coupling.
As long as the charge carriers (Mn $e_g$ holes) are mobile they
mediate a ferromagnetic interaction between the localised $S=3/2$
spins formed by the $t_{2g}$ electrons of the manganese. If the
electron lattice interaction is strong enough small polarons can arise
and the effective mass of the holes is increased by the lattice
distortion accompanying its motion through the crystal.  Due to this
reduction of the mobility the ferromagnetic double-exchange may break
down completely, but, as indicated by the mentioned experiments, this
happens in a spatially inhomogeneous way. We model this feature by
assuming a coexistence of a ferromagnetic metallic volume fraction,
described mainly by the mean-field theory of double-exchange by Kubo
and Ohata~\cite{KO72a}, and of a polaronic insulating volume fraction,
described by an exponentially narrow band and a paramagnetic spin
background.

\subsection{Ferromagnetic metallic component}
\begin{figure}[tb]
  \begin{center}
    \includegraphics[width=\linewidth]{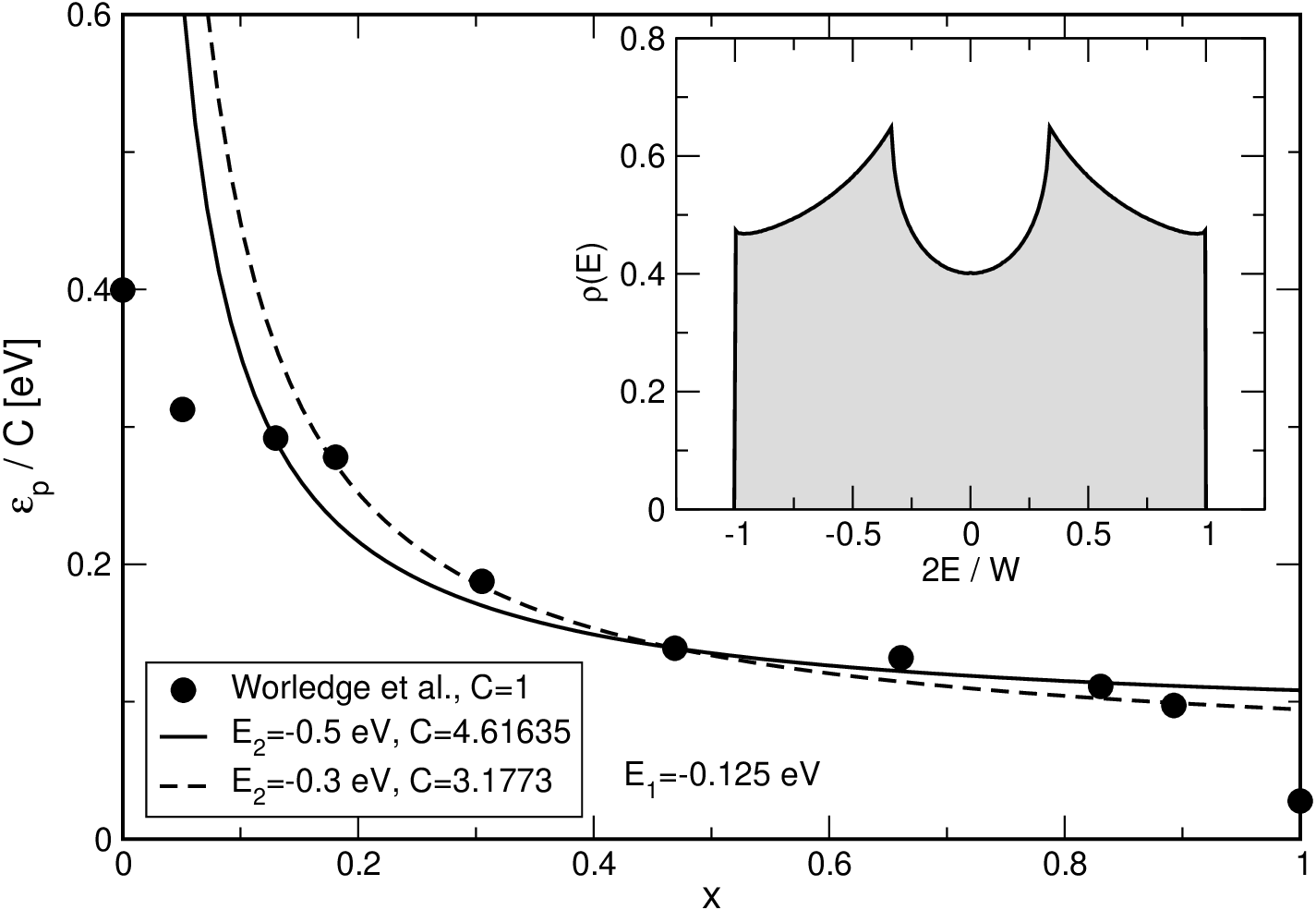}
  \end{center}
  \caption{Doping dependence of the polaronic binding energy: we
    compare experimental data~\cite{WMG98} from conductivity
    measurements (dots) with the $\epsilon_p$ from our ansatz (lines).
    Inset: Density of states used for the ferromagnetic metallic
    component.}\label{figepdos}
\end{figure}
Within mean-field theory the quantum double-exchange matrix
element~\cite{AH55} $\tilde t = t (S_T + 1/2)/(2S+1)$ for the hopping
of an itinerant charge carrier between two neighbouring sites is
averaged over all amplitudes and directions of the total bond spin
$S_T$, which is assumed to be placed in an inner Weiss field
$\lambda=\beta g \mu_B H_{\text{eff}}^z$ and an optional external
magnetic field $\lambda^{\text{ext}} = \beta g\mu_B H^z_{\text{ext}}$.
Omitting orbital degrees of freedom the resulting Hamiltonian
describes free fermions in an effective band of width
\begin{equation}\label{bandwidth}
  W = \gamma_S[S(\lambda+\lambda^{\text{ext}})]\, W_0\,,
\end{equation}
where $W_0$ denotes the bare band width and the field dependent
prefactor is given by~\cite{KO72a}
\begin{equation}\label{defgamma}
  \gamma_{S}[z] = \tfrac{S+1}{2S+1}
  + \tfrac{S}{2S+1}\coth\left(\tfrac{S+1}{S}z\right)B_S[z]
\end{equation}
with the Brillouin function
\begin{equation}\label{brillouin}
  B_S[z] =
  \frac{1}{2S}\left[(2S+1)\coth\tfrac{(2S+1)z}{2S}-\coth\tfrac{z}{2S}\right]\,.
\end{equation}
To improve the above approximation of Kubo and Ohata~\cite{KO72a}, we
account for the orbitally anisotropic hopping~\cite{An59,KK72}, which
follows from the perovskite structure,
\begin{equation}\label{tabxy}
  t_{\alpha\beta}^{x/y} =
  \frac{t}{4}\begin{bmatrix}
    1 & \mp\sqrt{3}\\
    \mp\sqrt{3} & 3
  \end{bmatrix}\,,
  \
  t_{\alpha\beta}^{z} =
  t \begin{bmatrix}
    1 & 0\\
    0 & 0
  \end{bmatrix}\,,
  \ 
  \alpha,\beta\in\{\theta,\varepsilon\}\,,
\end{equation}
and for the strong on-site Coulomb interaction. In a mean-field sense
both is achieved~\cite{WLF01b} by working with the averaged density of
states
\begin{equation}
  \varrho(E)=\frac{1}{2}(\varrho_{+}(E)+\varrho_{-}(E))\,,
\end{equation}
shown in the inset of Figure~\ref{figepdos}. The densities
$\varrho_\zeta(E)$ belong to the two bands ($\zeta=\pm 1$),
\begin{equation}
  \epsilon_{{\bf k},\zeta} = -t\Big(\sum_\delta \cos k_\delta
  +\zeta \sqrt{\tfrac{1}{4}\smash[b]{
      \sum_{\delta,\delta^\prime}(\cos k_{\delta}-\cos
      k_{\delta^\prime})^2}}\;\Bigl)\,,
\end{equation}
resulting from nearest neighbour matrix elements of
Equation~\eqref{tabxy} with $t=W/6$, Equation~\eqref{bandwidth}.

We assume this non-interacting fermion model to be valid within the
ferromagnetic volume fraction
\begin{equation}
  p^{(f)} = \frac{N^{(f)}}{N}
\end{equation}
of the sample. Of course $p^{(f)}$ is temperature and doping
dependent. Its actual value is determined self-consistently through
the equations given in Section~\ref{thermorel}.

\subsection{Polaronic insulating component}
In the remaining, polaronic part of the sample,
\begin{equation}
  p^{(p)} = \frac{N^{(p)}}{N} = 1 - p^{(f)}\,,
\end{equation}
we assume all charge carriers to be self-trapped small polarons, i.e.,
their kinetic energy is exponentially suppressed and the band centre
is shifted by a polaron binding energy $\epsilon_p$. To keep our model
as simple as possible, we completely neglect the polaronic band width
and consider only a dispersion-less level located at $\epsilon_p$. In
our previous work~\cite{WLF01b}, using certain energy arguments, we
motivated a doping dependence of $\epsilon_p$ of the form
\begin{equation}\label{defep}
  \epsilon_p = \left(x^{-1}-1\right) E_1 + E_2\,.
\end{equation}
Here $E_1$ and $E_2$ are effective model parameters describing the
anti-Jahn-Teller effect and the usual polaron binding energy,
respectively. If we discard the condition of equally charged metallic
and insulating volume fractions, in a strict sense the old derivation
is no longer valid. However, the comparison of the ansatz in
Equation~\eqref{defep} with experimental data~\cite{WMG98} for the
polaronic binding energy, which can be extracted from the temperature
and doping dependence of the resistivity in the high-temperature
paramagnetic phase, yields surprisingly good agreement over a wide doping
range. Figure~\ref{figepdos} illustrates that with our choice of
parameters $\{E_1,E_2\}$ the functional form of $\epsilon_p$ matches
the real data quite well, as long as $x$ is not too small. In absolute
values the curves disagree by a constant factor of the order $3$ to
$5$, which is sufficient for a mean-field type theory.

\section{Thermodynamic stability conditions}\label{thermorel}
Based on the above assumptions for the two coexisting components of
the low-temperature phase of the manganites we are now in the position
to formulate equilibrium conditions. The essential change to our
previous work concerns the assumption of equal pressure for carriers
in the metallic and the insulating region, where the pressures are
obtained from
\begin{align}
  \pi^{(f)} &= \frac{1}{\beta}\int \varrho(E)\, 
  \log(1+e^{\beta(\mu-E)})\, dE\,,\\
  \pi^{(p)} &= \frac{1}{\beta}\int \delta(E-\epsilon_p) \,
  \log(1+e^{\beta(\mu-E)})\, dE\,,
\end{align}
with $\beta = 1/(k_B T)$. For each value of the inner field $\lambda$
the equation
\begin{equation}\label{equpress}
  \pi^{(f)} = \pi^{(p)} =: \pi_{\text{eq}}
\end{equation}
then defines the chemical potential $\mu$, which is required to be
equal for both components. Given $\mu$ the resulting carrier
concentrations
\begin{align}
  x^{(f)} &= \int \frac{\varrho(E)}{e^{\beta(E-\mu)}+1}\, dE\\
  x^{(p)} &= \int \frac{\delta(E-\epsilon_p)}{e^{\beta(E-\mu)}+1}\, dE
\end{align}
in the coexisting regions define the two volume fractions by the
equations
\begin{equation}\label{doping}
  x = p^{(f)} x^{(f)} + p^{(p)} x^{(p)}
  \quad\text{and}\quad
  p^{(f)} + p^{(p)} = 1\,,
\end{equation}
which ensure the correct overall doping. Based on this set of
equations we are able to calculate the free energy per site for the
whole system,
\begin{equation}
  f = x\mu - \pi_{\text{eq}} + f^{(s)}\,,
\end{equation}
where
\begin{align}
  f^{(s)} & = \frac{1}{\beta}\Big\{
  x \Big[ p^{(f)} \big(\lambda S B_{S}[S(\lambda+\lambda^{\text{ext}})]\\
  & \qquad - \log \nu_{S}[S(\lambda+\lambda^{\text{ext}})]\big)
  - p^{(p)} \log \nu_{S}[S\lambda^{\text{ext}}]\Big]\notag\\
  & + (1-x) \Big[ p^{(f)} \big(\lambda \bar{S}
  B_{\bar{S}}[\bar{S}(\lambda+\lambda^{\text{ext}})]\notag\\
  & \qquad - \log
  \nu_{\bar{S}}[\bar{S}(\lambda+\lambda^{\text{ext}})]\big) - p^{(p)}
  \log \nu_{\bar{S}}[\bar{S}\lambda^{\text{ext}}]\Big]\Big\}\notag
\end{align}
denotes the spin part of the free energy ($\bar{S} = S+\frac{1}{2} =
2$) and
\begin{equation}
  \nu_{S}[z] = \sinh(z)\coth\left(\tfrac{z}{2S}\right)+\cosh(z)
\end{equation}
is the spin partition function.

The set of equilibrium conditions is closed by the requirement that
$\lambda$ is chosen such that the free energy is minimal. The same
is required, if the above equations have more than a single
solution.  A finite value of $\lambda$ corresponds to a
ferromagnetically ordered metallic component, and if in addition the
corresponding volume fraction $p^{(f)}$ is nonzero the magnetisation
of the whole sample is given by
\begin{multline}
  m = (1-x)\left[p^{(f)} \bar{S} B_{\bar
      S}[\bar{S}(\lambda+\lambda^{\text{ext}})]
    + p^{(p)}\bar{S} B_{\bar S}[\bar{S}\lambda^{\text{ext}}]\right] \\
  + x\left[p^{(f)} S B_{S}[S(\lambda+\lambda^{\text{ext}})] + p^{(p)}S
    B_{S}[S\lambda^{\text{ext}}]\right]\,.
\end{multline}

\section{DC conductivity}
The above thermodynamic relations define the phase boundary between the
ferromagnetic metallic and the paramagnetic insulating phase and
explain the behaviour of the magnetisation and of the volume fractions
of the different components. However, they do not contain any
information about the resistivity of the system.

In the past, approaches which are based on the percolative mixing of
regions with different macroscopic resistivities have been
successfully used to fit experimental data~\cite{Jaea99,LS96,Beea02}.
Here we follow a similar path to model the resistivity $\rho$ of our
mixed-phase system. Namely, we assume that the resistivity of the
metallic component is proportional to the expression
\begin{align}
  \rho_S[z] & = \frac{g_S[z] - \gamma_S[z]^2}{\gamma_S[z]^2}\\
  g_S[z] & = \frac{S
    B_S[z]}{(2S+1)^2}\left[(2S+2)\coth\frac{(S+1)z}{S} -
    \coth\frac{z}{2S}\right]\notag\\
  & + \frac{S+1}{2S+1} \,,
\end{align}
derived by Kubo and Ohata~\cite{KO72a}, which associates $\rho$ with
the fluctuation of the double-exchange matrix element caused by the
thermal spin disorder. The resistivity of the insulating component is
assumed to match the resistivity of the high-temperature phase, which
in experiment is well fit by the activated hopping of
small-polarons~\cite{EH69,JSRTHC96,WMG98}. Hence, the resistivities of
the two components are given by,
\begin{align}
  \rho^{(f)} & = \frac{B}{x^{(f)}}
  \left(\rho_S[S(\lambda+\lambda^{\text{ext}})]+\rho_{\text{min}}\right)\,,\\
  \rho^{(p)} & = \frac{A}{\beta x^{(p)}}\ \rho_S[S\lambda^{\text{ext}}]
  \ e^{-\beta \epsilon_p}\,,
\end{align}
where the prefactors $A$ and $B$ as well as the cut-off
$\rho_{\text{min}}$ are free model parameters which could be
estimated from experimental data. 

The resistivity of the whole sample, which may consist of an
inhomogeneous mixture of both components, is calculated by assuming a
random resistor network. More precisely, we choose nodes from a cubic
lattice which belong to the metallic component with probability
$p^{(f)}$ and to the polaronic component with probability $p^{(p)}$.
Each of these nodes, which represent macroscopic regions of the
sample, is connected to its neighbours with resistors of
magnitude $\rho^{(f)}$ or $\rho^{(p)}$, respectively. The resistivity
of the network then yields a reasonable approximation for the
resistivity of the inhomogeneous ferromagnetic metallic phase of the
manganites. The percolative nature of this model, particularly in the
vicinity of the phase transition, makes the system very sensitive to
small changes in temperature, doping and external magnetic field.

\section{Results}
\begin{figure}[tb]
  \begin{center}
    \includegraphics[width=\linewidth]{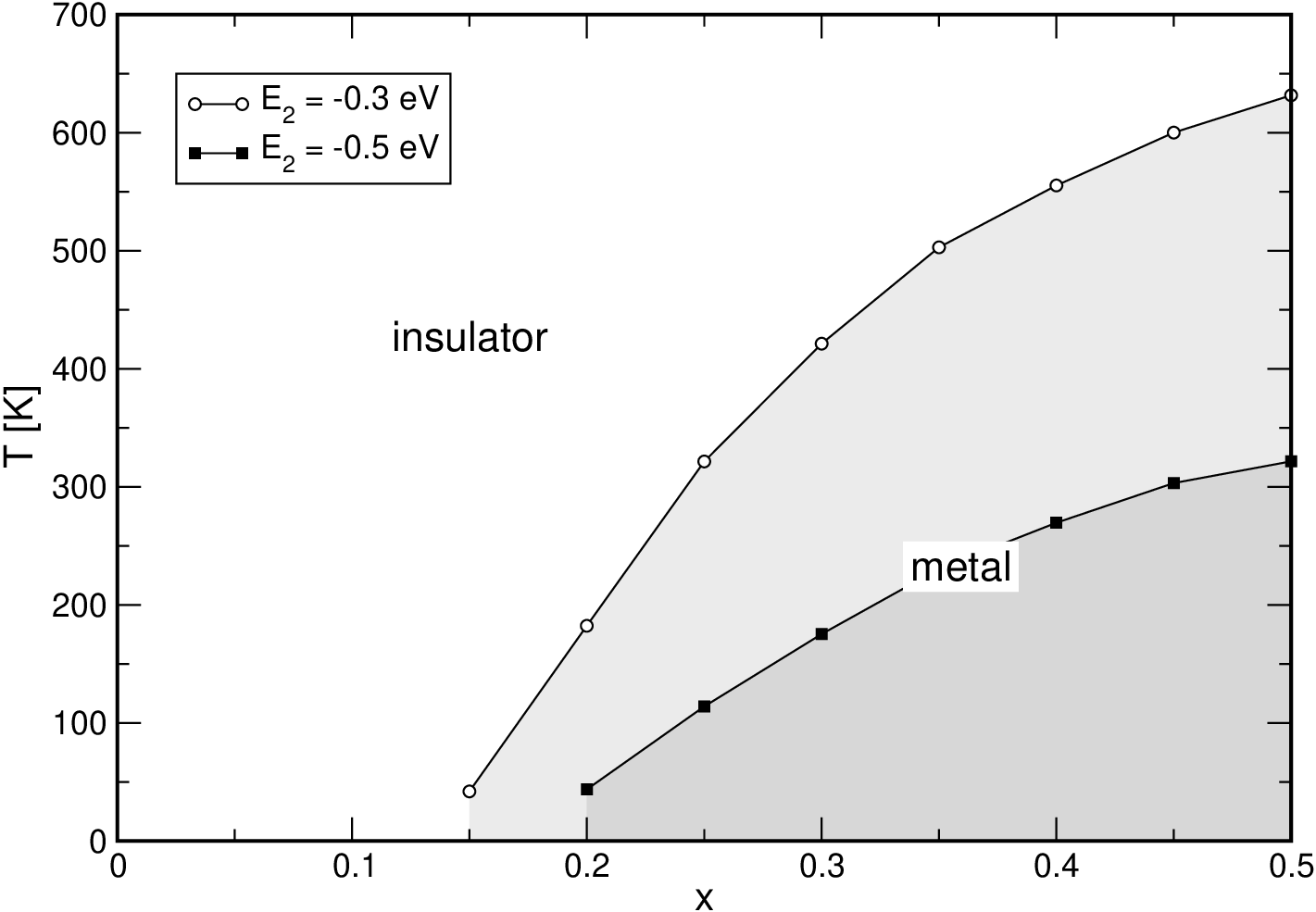}
  \end{center}
  \caption{$x$-$T$ phase diagram of the mixed-phase model obtained for
    $W_0 = 2.4$~eV, $E_1 = -0.125$~eV and the two $E_2$ values
    $-0.3$~eV and $-0.5$~eV.}\label{figphase}
\end{figure}
\begin{figure}[tb]
  \begin{center}
    \includegraphics[width=\linewidth]{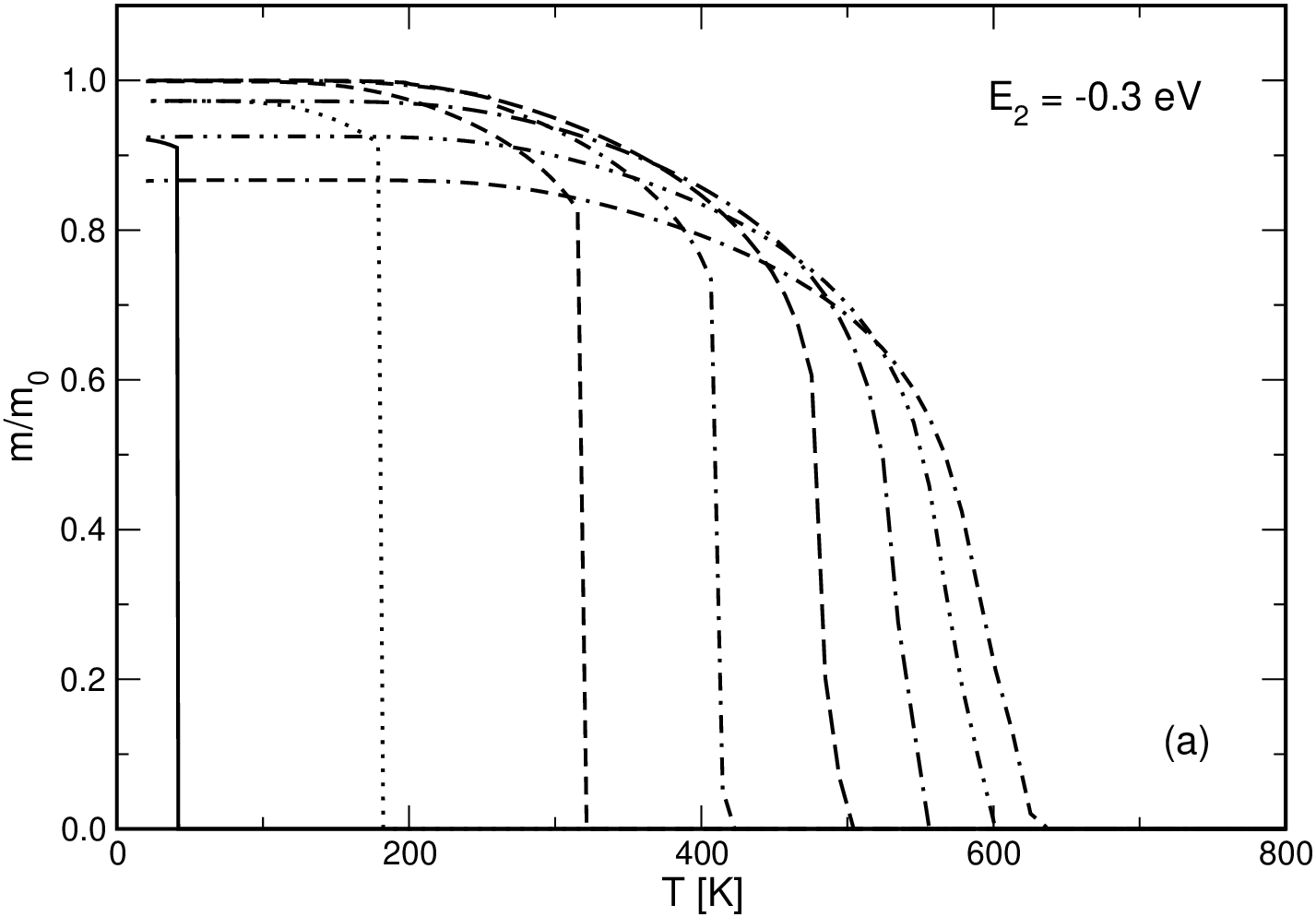}
    \includegraphics[width=\linewidth]{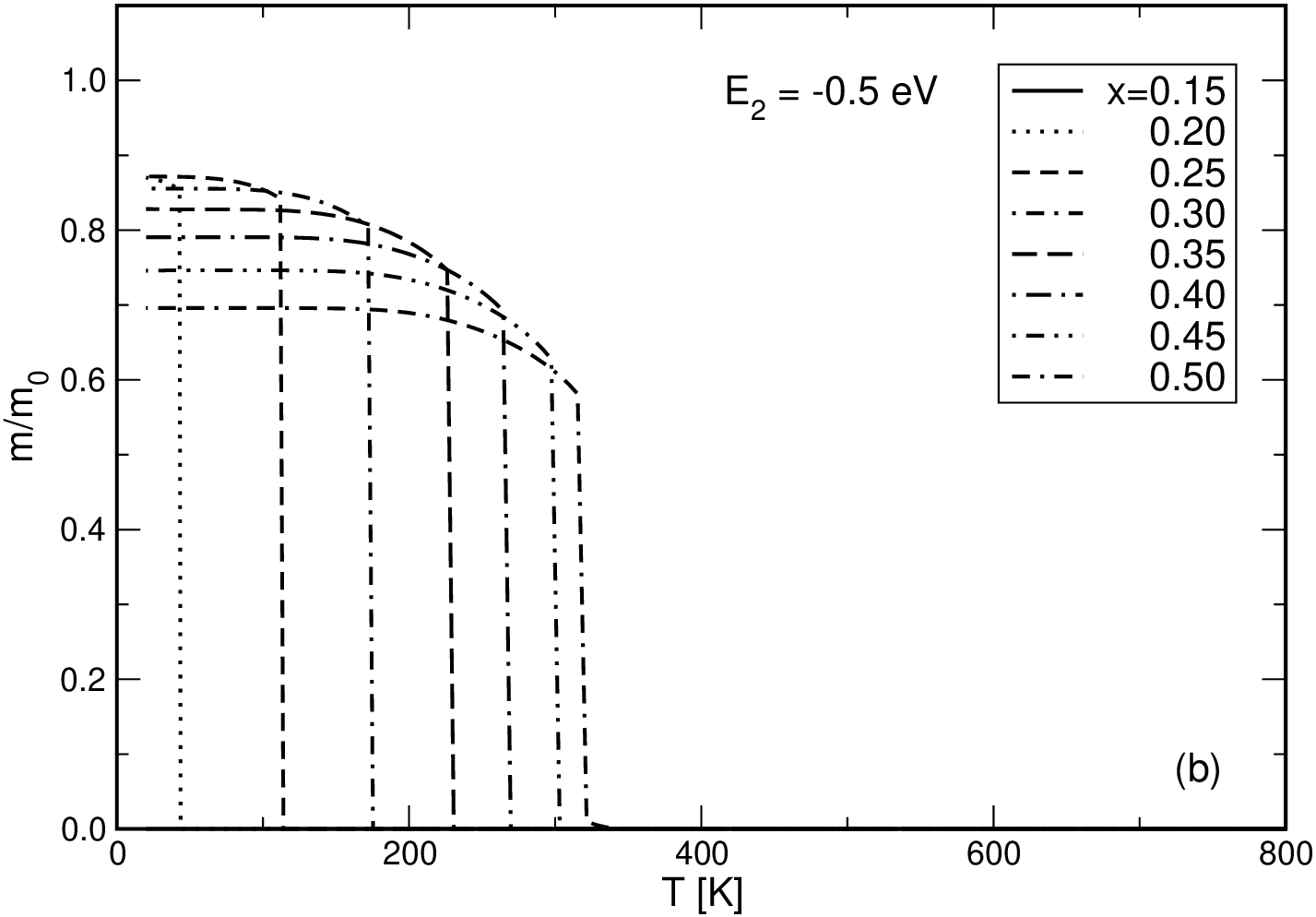}
  \end{center}
  \caption{Temperature dependence of the magnetisation
    $m$ for the two parameter sets and the indicated doping
    levels. Normalisation: $m_0=\bar S - x/2$.}\label{figmag}
\end{figure}
The numerical solution of the self-consistency equations is rather
straightforward. However, some care is recommended, if there are
multiple solutions for the equations~\eqref{equpress}--\eqref{doping}.
To give an example, we set the bare band-width and the Jahn-Teller
energy equal to $W_0=2.4$~eV and $E_1=-0.125$~eV, respectively, and
consider two typical values for the polaronic binding energy,
$E_2=-0.3$~eV and $-0.5$~eV. This choice results in the phase
boundaries displayed in Figure~\ref{figphase}.  Without an external
magnetic field, for each doping $x$, the transition is defined by the
critical temperature $T_C$ for which the magnetisation $m$ of the
sample vanishes. The model yields reasonable values for both, the
transition temperatures and the critical doping $x_c$ at $T=0$.

Figure~\ref{figmag} shows the magnetisation $m$ as a function of
temperature and doping.  Clearly, the order of the phase transition
depends on both, the polaronic parameters $\{E_1, E_2\}$ {\em and} the
doping~$x$. Higher transition temperatures usually correspond to a
continuous, second order transition, whereas otherwise the transition
is first order.  Similar behaviour was also found for the real
materials~\cite{MRRVL99,Kiea02}.

\begin{figure}[tb]
  \begin{center}
    \includegraphics[width=\linewidth]{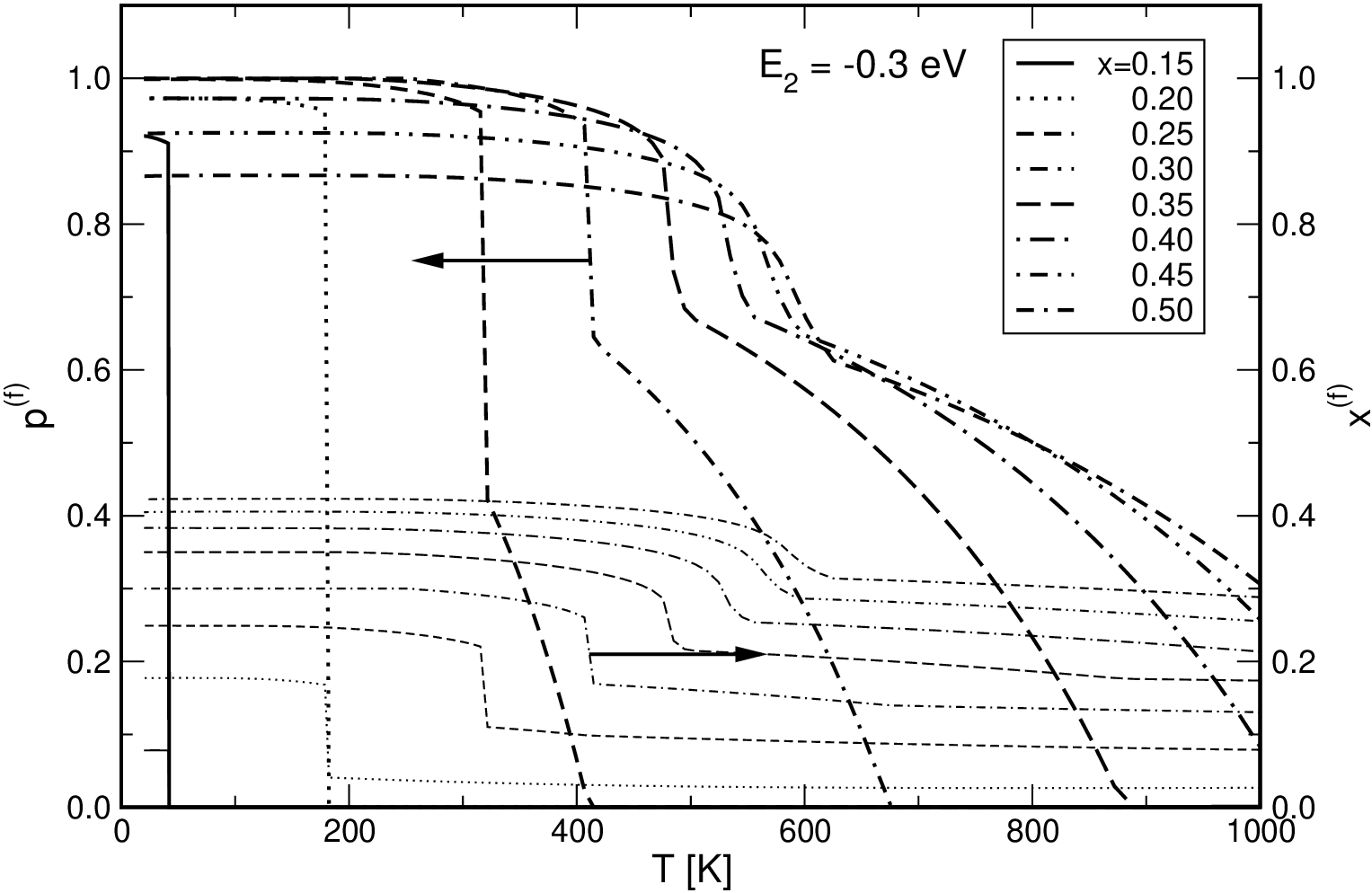}
  \end{center}
  \caption{Ferromagnetic volume fraction $p^{(f)}$ (bold lines) and
    corresponding carrier concentrations $x^{(f)}$ (thin
    lines) calculated with $E_2=-0.3$~eV.}\label{figpfxf}
\end{figure}
The inhomogeneous nature of the ferromagnetic and the paramagnetic
phase becomes evident from Figure~\ref{figpfxf}, where we show the
dependence of the ferromagnetic volume fraction $p^{(f)}$ and of the
corresponding carrier concentration $x^{(f)}$ on temperature and
doping. In the case of small $T_C$ the sample is never completely
metallic, i.e., a small polaronic insulating volume fraction is
present even at lowest temperatures. On the other hand, for higher
$T_C$ a finite metallic volume fraction exists also above $T_C$. Since
it is usually smaller than the percolation threshold the sample
remains insulating. In addition the carrier concentration within the
metallic component is noticeable reduced above $T_C$.  In the case of
a second order transition $p^{(f)}$ and $x^{(f)}$ also decrease
continuously with temperature. The presence of a metallic,
double-exchange driven component above $T_C$ can be related to the
experimentally observed ferromagnetic correlations, sometimes
interpreted as ferromagnetic clusters or magnetic
polarons~\cite{DTea96,CNIFGJG99,CRMC98,DMS01}.

\begin{figure}[tb]
  \begin{center}
    \includegraphics[width=\linewidth]{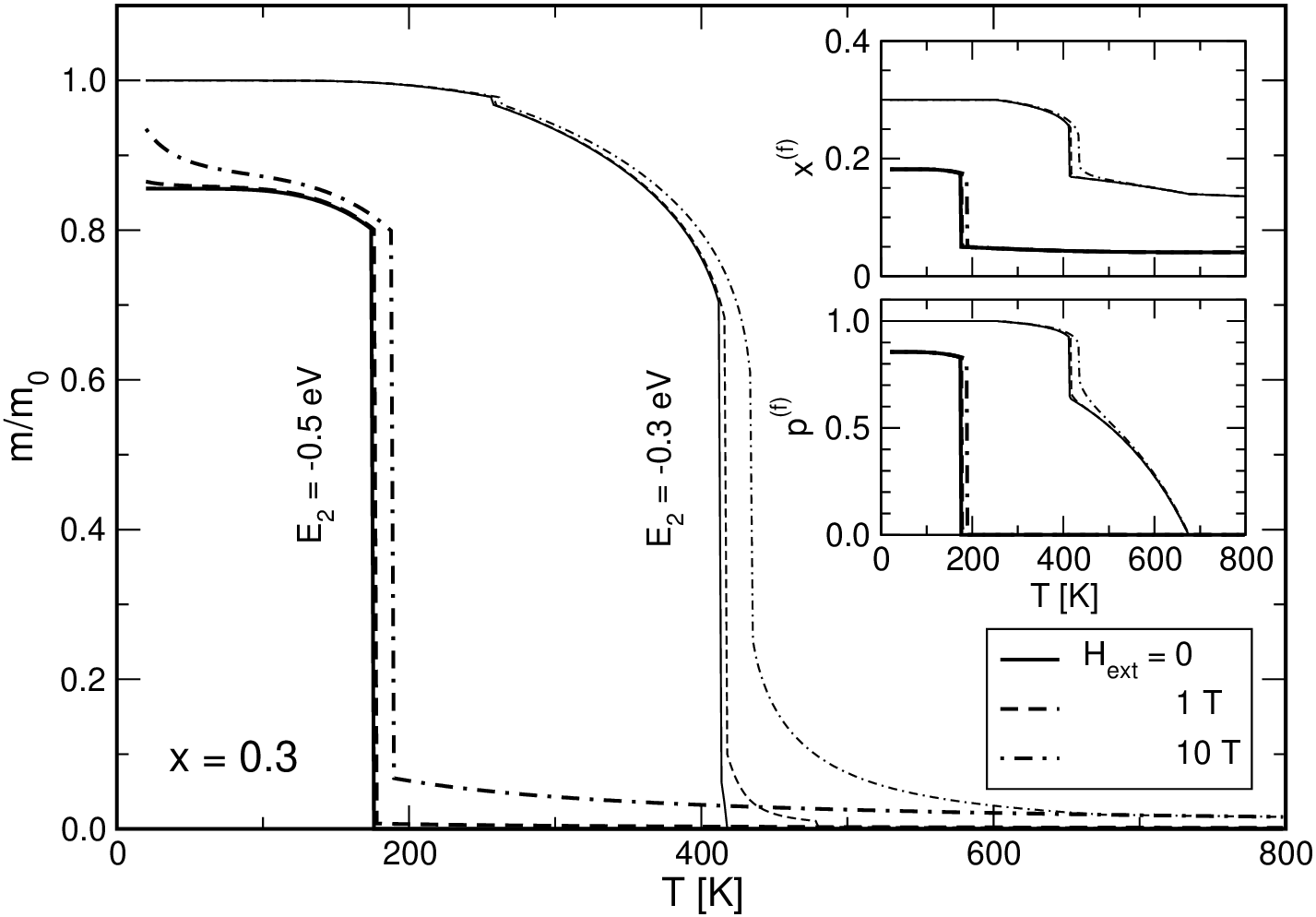}
  \end{center}
  \caption{Influence of an external magnetic field $H^{z}_{\text{ext}}$
    on the magnetisation $m$ (main panel), the volume fraction
    $p^{(f)}$ and the carrier concentration $x^{(f)}$ (insets).}
  \label{figtfhext}
\end{figure}
The sensitivity of the phase transition and of all related quantities
to an external magnetic field is illustrated in Figure~\ref{figtfhext}.
There we show the magnetisation $m$ together with the volume fraction
$p^{(f)}$ and the carrier concentration $x^{(f)}$ of the ferromagnetic
metallic component for the two considered parameter sets and a doping
level of $x=0.3$. Clearly, even for moderate field strength the
critical temperature is shifted by a few degrees and in particular the
volume fraction $p^{(f)}$ changes noticeable around $T_C$. Of course,
the latter has an important influence on the conductivity of the
system.

\begin{figure}[tb]
  \begin{center}
    \includegraphics[width=\linewidth]{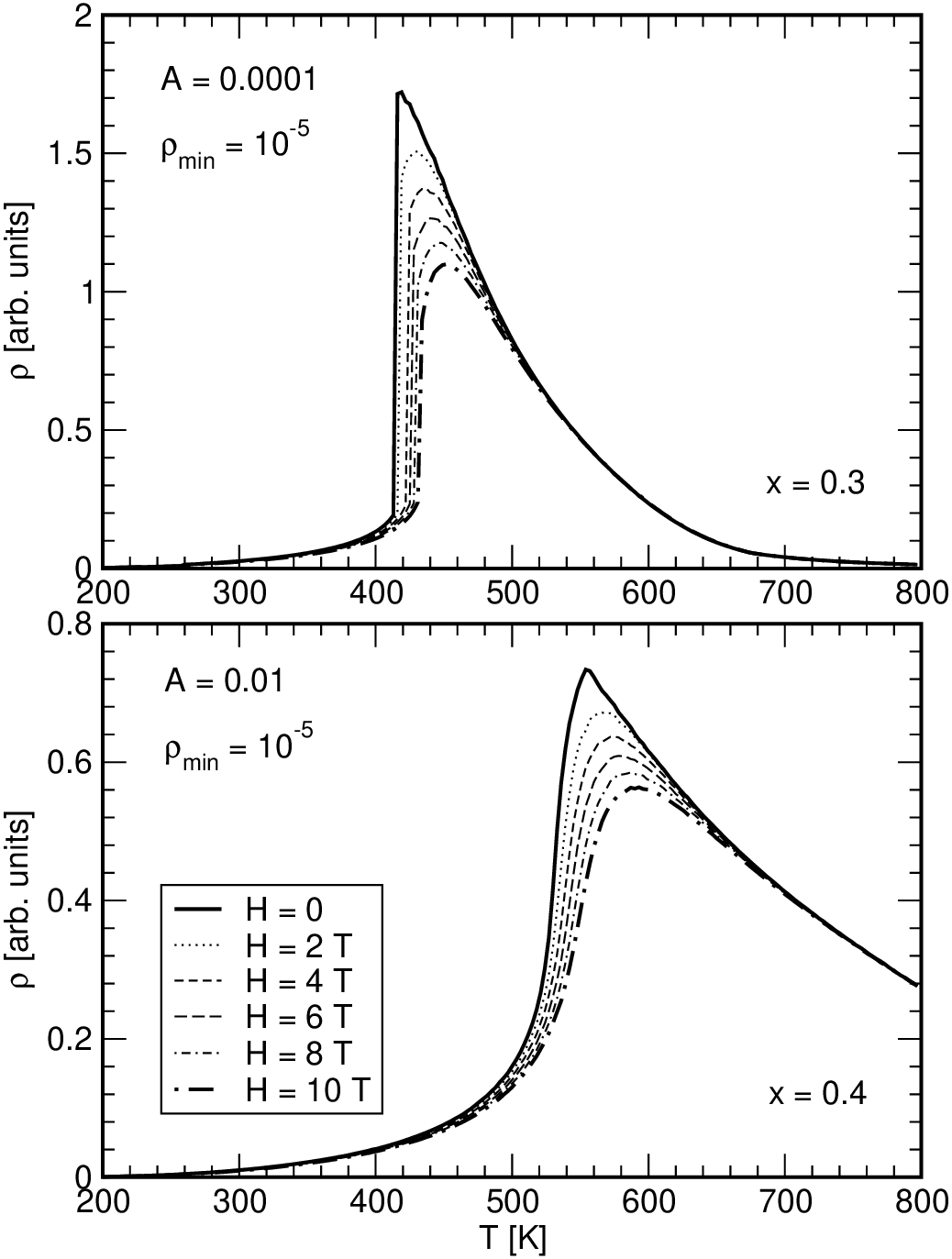}
  \end{center}
  \caption{Resistivity $\rho$ under the influence of an external
    magnetic field $H^{z}_{\text{ext}}$ for doping $x=0.3$ and $0.4$
    and the parameter set $\{W_0,E_1,E_2\} = \{2.4,-0.125,-0.3\}$~eV.}
  \label{figrho}
\end{figure}
Inserting the volume fractions and carrier concentrations from the
mixed-phase model into the ansatz for the DC conductivity we obtain the
resistivities $\rho$ shown in Figure~\ref{figrho}. Since we are mainly
interested in the general features of $\rho$, we set $B=1$ and use the
specified values for $A$ and $\rho_{\text{min}}$. Depending on the
order of the phase transition $\rho$ shows a sharp jump or a
continuous increase close to $T_C$. This behaviour of the resistivity
originates to a large degree from the changing volume fraction of the
metallic component, which can cross the percolation threshold.
However, the conductivity of the component itself as well as its
carrier concentration strongly affect $\rho$ for $T<T_C$. An external
magnetic field causes a reasonable suppression of $\rho$, i.e., a
noticeable negative magnetoresistance.  Compared to the real compounds
the calculated effect is a bit weaker. Nevertheless, in view of the
rather simple model for the conductivity the agreement is quite
satisfactory.  Probably, more involved assumptions for the
resistivities of the two different components could improve the
magnetoresistance data. At present the conductivity of the metallic
component is controlled only by spin fluctuations in the
double-exchange hopping, which could be increased by
the weak anti-ferromagnetic interactions active in the manganites.
In addition, other scattering mechanisms could play a role.  The
adiabatic small-polaron approximation used for the polaronic
insulating volume fraction may be questionable as well, since the
phonon modes involved in the electron-lattice coupling do not have
small enough frequency. Another potential improvement concerns the
site percolation model used to construct the random resistor network.
As was pointed out recently, an approach that is based on correlated
percolation could be more appropriate~\cite{KK02}. An affinity to the
formation of larger regions of the same type would naturally affect
the resistivity of the system and its response to an external field.

\section{Conclusions}
In summary, we have proposed a phenomenological model for the
ferromagnetic metallic phase of doped colossal magnetoresistive
manganites, which is based on the coexistence of a double-exchange
driven metallic component and a polaronic insulating component. Using
modified equilibrium conditions and adding a percolative ansatz for
the resistivity of the mixed phase we have substantially improved
previous work~\cite{WLF01b}. With realistic parameters for the
electronic band-width, the Jahn-Teller splitting and the polaronic
binding energy our approach yields reasonable data for the phase
boundary of the ferromagnetic metallic phase and correctly predicts
the existence of a finite polaronic volume fraction well below the
critical temperature $T_C$. The model shows a manifest sensitivity to
external magnetic fields, including a large negative magnetoresistance
close to $T_C$.

We thank F. G\"ohmann and N. Shannon for valuable discussions and
acknowledge financial support by the Deutsche Forschungsgemeinschaft
and the Czech Academy of Sciences under grant No. 436 TSE 113/33/\mbox{0-2}.

\end{document}